# Near-zero-field microwave-free magnetometry with nitrogen-vacancy centers in nanodiamonds

Omkar Dhungel[1,2], Mariusz Mrózek[3,*], Till Lenz[1,2], Viktor Ivády[4,5,6], Adam Gali[7,8,9], Arne Wickenbrock[1,2], Dmitry Budker[1,2,10], Wojciech Gawlik[3] and Adam M. Wojciechowski[3]

[1]Helmholtz-Institut Mainz, GSI Helmholtzzentrum für Schwerionenforschung GmbH, 55128 Mainz, Germany
[2]Johannes Gutenberg-Universität Mainz, 55128 Mainz, Germany
[3]Jagiellonian University, Faculty of Physics, Astronomy and Applied Computer Science, Lojasiewicza St. 11, 30-348 Krakow, Poland
[4]Department of Physics of Complex Systems, ELTE Eötvös Loránd University, Egyetem tér 1-3, H-1053 Budapest, Hungary
[5]MTA-ELTE Lendület "Momentum" NewQubit Research Group, Pázmány Péter, Sétány 1/A, 1117 Budapest, Hungary
[6]Department of Physics, Chemistry and Biology, Linköping University, 581 83 Linköping, Sweden
[7]Wigner Research Centre for Physics, P.O. Box 49, H-1525 Budapest, Hungary
[8]Budapest University of Technology and Economics, Institute of Physics, Department of Atomic Physics, Műegyetem rakpart 3., 1111 Budapest, Hungary
[9]MTA-WFK Lendület "Momentum" Semiconductor Nanostructures Research Group, P.O. Box 49, H-1525 Budapest, Hungary
[10]Department of Physics, University of California, Berkeley, California 94720-300, USA

We study the fluorescence of nanodiamond ensembles as a function of static external magnetic field and observe characteristic dip features close to the zero field with potential for magnetometry applications. We analyze the dependence of the feature's width and the contrast of the feature on the size of the diamond (in the range 30 nm–3000 nm) and on the strength of a bias magnetic field applied transversely to the field being scanned. We also perform optically detected magnetic resonance (ODMR) measurements to quantify the strain splitting of the zero-field ODMR resonance across various nanodiamond sizes and compare it with the width and contrast measurements of the zero-field fluorescence features for both nanodiamonds and bulk samples. The observed properties provide compelling evidence of cross-relaxation effects in the NV system occurring close to zero magnetic fields. Finally, the potential of this technique for use in practical magnetometry is discussed.

## Introduction

The negatively charged nitrogen-vacancy (NV⁻) center in diamond has interesting optical and spin properties, making it useful for developing sensors for magnetic and electric fields, temperature, and pressure[1,2,3,4,5]. NV-center-containing nanodiamonds are non-toxic and photostable and can be easily functionalized, so that they can be used as fluorescent markers and sensors in biological materials[6,7].

A widely employed research and magnetometry technique for NV centers is optically detected magnetic resonance . It is an electron spin-resonance technique by which the spin state of a paramagnetic structural defect in wide band-gap semiconductors may be optically pumped for spin initialization and readout. To measure weak magnetic fields, this method usually requires a bias field to lift the degeneracy between the $m_s = \pm 1$ sublevels of the triplet ground state of the defect. Due to the applied high external magnetic fields and/or the application of microwave fields, the ODMR technique may be problematic for studying sensitive samples where either the external magnetic field or the microwave irradiation might disturb the system. To solve this problem, various measurement techniques have been developed, for example, microwave-free magnetometry[8] based on the level anti-crossing in a high magnetic field (102.4 mT for the NV center) or using circularly polarized microwave fields to individually address transitions to the

---

$m_s = +1$ or $m_s = -1$ states[9]. Recently, low-field microwave-free magnetometry has been proposed[10] relying on the measurement of the position of the cross-relaxation features[11,12].

In this work, we propose to utilize ensembles of nanodiamonds for low-field MW-free magnetometry and perform a detailed study of the zero-field fluorescence features in nanodiamonds of different diameters containing NV centers. We demonstrate that the observed zero-field features in the form of dips in fluorescence intensity correlate well with the changes in the longitudinal relaxation rate at close to zero magnetic fields. We numerically predict and experimentally verify the observed cross-relaxation feature as a function of a field applied in one direction with a bias field applied in a transverse direction. These results are also compared with those obtained for bulk NV diamonds. In this way, the article extends on the research presented in Ref. [13] for bulk samples.

**Experimental setup**

Measurements were performed using a confocal microscope arrangement shown schematically in Fig. 1(a). A microscope objective (Motic 10×, NA = 0.25) was used to focus the exciting 532 nm laser beam (Sprout-G) with stabilized light power at a spot of size 4 μm$^2$. A dichroic mirror (Thorlabs DMLP567), long-pass filter (Thorlabs FEL0600), and avalanche photodiode (Thorlabs APD440A2) allowed the detection of light in a wavelength range of approximately 600–850 nm, and the fluorescence signal was processed with an oscilloscope. A programmable power supply (HAMEG 4040) was connected to the XY coils (X coils 16.94 G/A, Y coils 7.34 G/A). The power supply was programmed with a current ramp comprising 110 points in the range of 0 to 3 A; the current sweep also triggered the scope. The ramped currents from the device were unidirectional, but their polarity was changed from positive to negative during the experiment. The dependence of the fluorescence signal on the applied current was averaged 512 times and recorded with a scope.

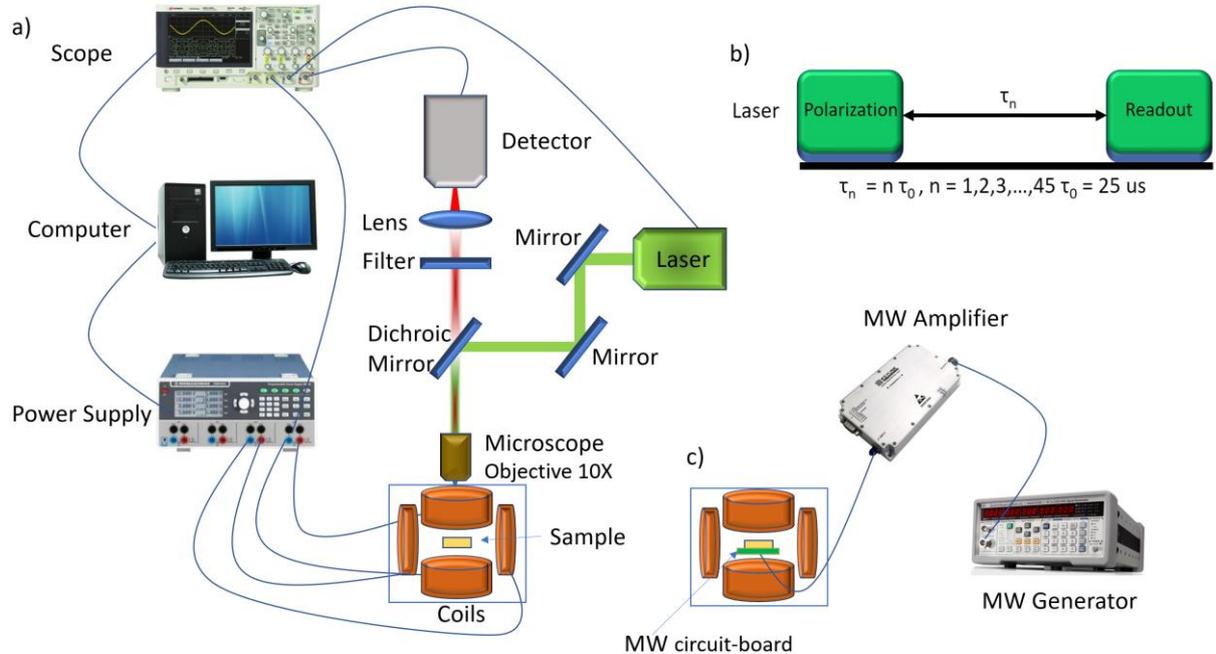

Figure 1 (a) Schematic diagram of the experimental setup. (b) Pulse sequence used for T$_1$ measurements. (c) Additional elements with MW circuit board used for ODMR strain measurements.

For strain measurement, the apparatus was supplemented with a MW part which consisted of a MW generator (SG386, SRS) and a MW amplifier (KB0842M47A, Keylink). The generator signal was fed to a MW "antenna" on a printed circuit board (PCB). We used a resonant PCB structure with a central opening (2 mm wide) that yielded a homogeneous MW power over the investigated sample surface. A detailed MW-circuit description is provided in earlier work[14,15]. Longitudinal relaxation times ($T_1$) were measured using the "relaxation-in-the-dark method"[16]. Figure 1(b) shows the sequence of the optical pulses used to measure $T_1$.

Table 1 provides details of the diamond samples used in the experiment.

| Sample | Manufacturer | NV concentration | Crystallographic orientation | Surface functionalization |
|---|---|---|---|---|
| ND 30 nm | Adamas Nanotechnologies | ~1.5 ppm | random | -COOH |
| ND 40 nm | Adamas Nanotechnologies | ~1.5 ppm | random | -COOH |
| ND 50 nm | Adamas Nanotechnologies | ~2 ppm | random | -COOH |
| ND 70 nm | Adamas Nanotechnologies | ~3 ppm | random | -COOH |
| ND 100 nm | Adamas Nanotechnologies | ~3 ppm | random | -COOH |
| ND 140 nm | Adamas Nanotechnologies | ~3 ppm | random | -COOH |
| ND 250 nm | Petr Cigler group | no information | random | - |
| ND 300 nm | Adamas Nanotechnologies | ~3 ppm | random | -COOH |
| ND 750 nm | Adamas Nanotechnologies | ~3.5 ppm | random | -COOH |
| ND 3000 nm | Adamas Nanotechnologies | ~3.5 ppm | random | -COOH |
| Bulk | dEYEmond GmbH | ~7 ppm | 100 | - |

Table 1. List of samples (nanodiamond and bulk) used in the experiments.

## Results

First, we unambiguously prove that the zero-field feature observed in the fluorescence signal is related to the enhanced relaxation rate of the electron spin states close to zero magnetic fields. To this end, we measured the magnetic-field dependence of the longitudinal relaxation time ($T_1$) and compared it with the magnetic-field dependence of the PL intensity. Previous research on nitrogen-vacancy color centers in diamond showed a strong dependence of $T_1$ on the temperature, direction, and strength of the magnetic field[12,17]. It was noticed that the value of the longitudinal relaxation time follows the fluorescence level, depending on the magnetic field[18]. In Figure 2, we present the results of measurements of NV relaxation times around zero magnetic fields (red points) for a bulk sample as a reference (a), and for 750 nm diamond (b) and 140 nm nanodiamond samples (c). The zero-field resonance shapes for the corresponding samples are shown in the background (dark-gray curves). It can be seen that the $T_1$ value closely follows the fluorescence level for the varying magnetic field. We regard this observation as direct evidence of the relaxation occurring at the zero-field feature.

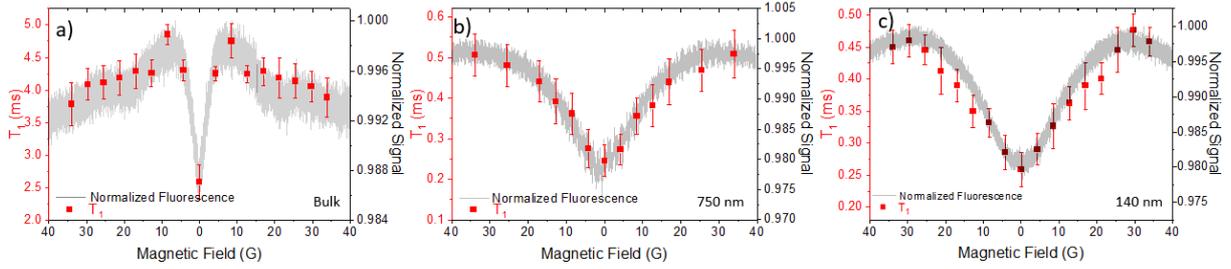

Figure 2 (a) Values of longitudinal relaxation times (red dots) compared with fluorescence intensity around zero-field resonance (gray curve) for a bulk sample, while figures (b) and (c) present analogous comparisons for 750-nm diamonds and 140-nm nanodiamonds, respectively.

Next, we explore the properties of the fluorescence feature close to zero magnetic fields in nanodiamonds with an estimated NV density of 1.5 to 3.5 ppm and various sizes. Table I details the source and main properties of the samples investigated. In such samples, we find that the width and contrast of the zero-field feature depend on the laser intensity.

Characterization of the zero-field feature, i.e. determination of its width (FWHM) and amplitude, was performed with a laser spot size of 4 µm². Figure 3 shows the zero-field features for (a) a bulk sample, (b) 1 µm nanodiamonds, and (c) 100 nm nanodiamonds recorded for various laser powers. The displayed signals that span the magnetic field range from -40 to +40 G were taken by stitching the curves recorded when the coil current was ramped both for positive and negative polarities.

*Laser-power dependence*

Figures 3(d) and 3(e) show the laser power dependences of the main characteristics of the central dip in the fluorescence intensity appearing at zero field: (d) the dip width and (e) its contrast. As can be seen, for low laser power, both parameters increase with the laser power, then both the width and contrast peak at approximately the same power, and then decrease for higher light powers. In the power dependence of the width, we initially observe a familiar power broadening, but for higher powers the opposite behavior is manifested, the *light narrowing*, most likely related to the effect discussed in Ref. [19]. The low contrast and width values for low laser powers are associated with the low efficiency of spin polarization of the NV color centers in nanodiamonds. Interestingly, the laser power dependences observed for nano- and bulk diamond peaks at comparable light powers of about 0.2 mW differ only in their amplitudes.

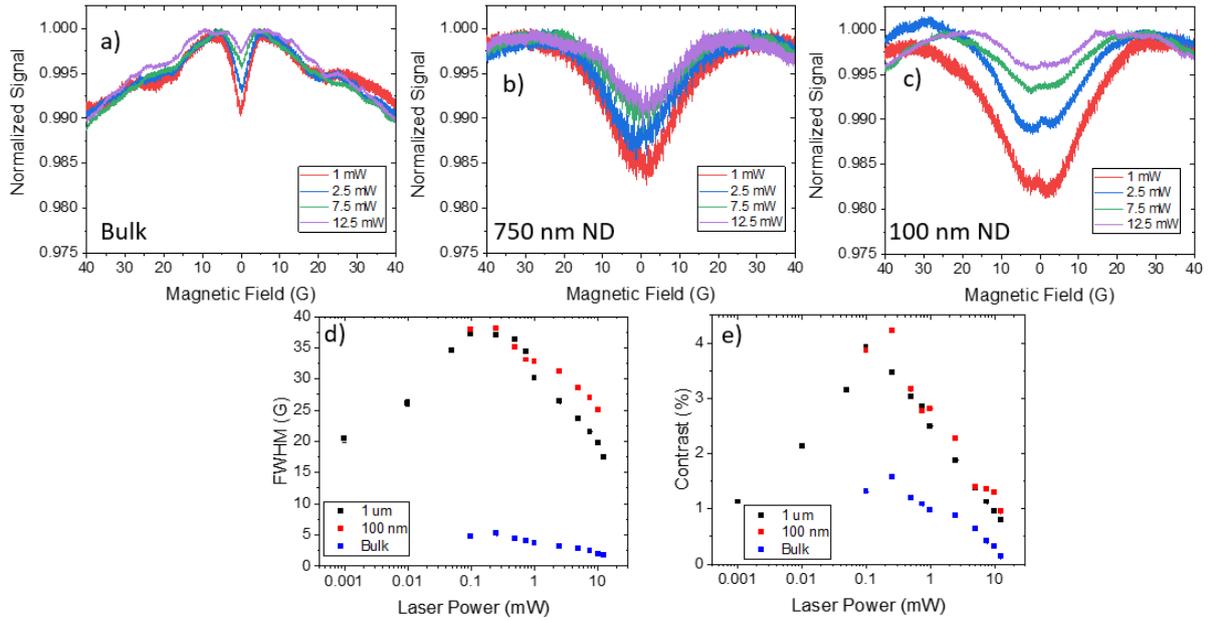

Figure 3 Zero-field resonances at various pumping laser powers: (a) a bulk sample, (b) 1 μm nanodiamonds, and (c) 100 nm nanodiamonds. Figure (d) shows the resonance width (FWHM) for different samples as a function of the laser power (semi-logarithmic scale); Figure (e) shows the resonance contrast for different samples vs. laser power (semi-logarithmic scale).

*Size dependence*

An interesting and important question is whether the average size of the nanodiamonds affects the width and contrast of the zero-field feature. We used several nanodiamond sizes between 30 nm and 3000 nm (from *Adamas Nanotechnologies, USA* and the Dr. Peter Cigler group, Institute of Organic Chemistry and Biochemistry of the Czech Academy of Sciences) with comparable NV concentrations as listed in Table 1. Figure 4 shows the size dependence of resonance contrast (a), width (b), and strain splitting of the ODMR signal (c) for different ND sizes. The gray-shadowed areas indicate the ranges of corresponding values measured with various bulk samples. All measurements were made with a constant laser power of 0.75 mW. The ND size-contrast relationship is shown in Fig. 4(a). For small ND sizes, the contrast initially increases until it is about four times higher than those obtained for a bulk diamond, and then it slowly decreases and approaches the values observed in the bulk sample. For the width data shown in Fig. 4(b), we observe a monotonous dependence, the width quickly decreases with the ND size and asymptotically approaches the results observed for bulk samples.

An important factor that affects the properties of the investigated fluorescence feature is the strain/electric field in the nanocrystal lattice, which changes with the diamond size. To study these effects, we performed ODMR measurements at a zero magnetic field. For this purpose, we slightly modified the setup by inserting a MW strip line (Fig.1(c)). Figure 4(c) shows the results obtained. The E splitting of the $|\pm\rangle$ states (determined by fitting the net ODMR resonance shape by two Lorentzian components) depends on the ND size similar to that of the width of the zero-field feature width.

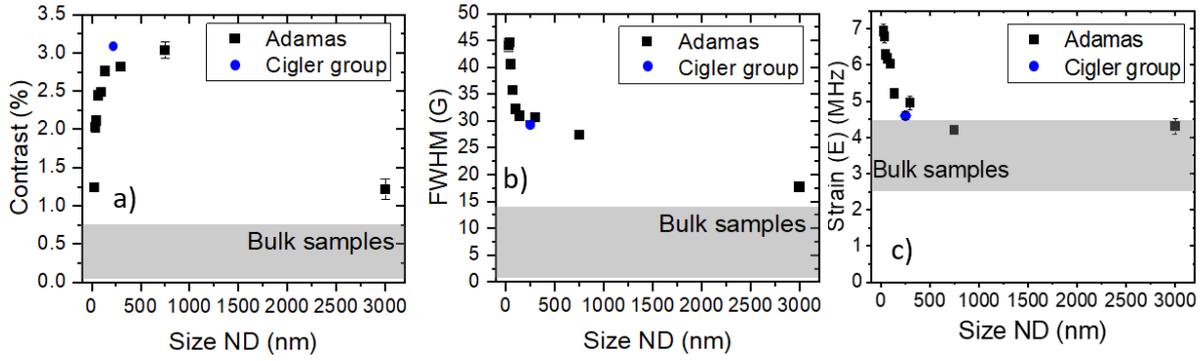

Figure 4 Size dependences of the properties of the zero-field resonance: (a) resonance contrast, (b) width, and size dependence of the strain splitting of the ODMR resonance in a zero magnetic field (c). Data taken with samples from Adamas are marked in black squares, and those from the Peter Cigler group are marked in blue circle. The gray-shadowed box indicates the range of values measured with various bulk samples.

*Effects of the transverse field*

The results described above were obtained with the longitudinal orientation of the magnetic field. However, we have also studied the effect of changing the magnetic field direction by applying a transverse magnetic field $B_{trans}$ in addition to the longitudinal one $B_{long}$. One known effect of the transverse field is the cross-relaxation that occurs when various eigenfrequencies in NV spectra become degenerate, which may happen for various crystallographic orientations and magnetic fields[12,13].

Figure 5 (a) shows plots of the experimentally recorded cross-relaxation features for different intensities of the transverse field with a bulk diamond. The bulk sample is cut in the crystallographic direction [100]. One observed effect of the cross-relaxation induced by a weak transverse field in the bulk sample is the broadening of the zero-field feature. For higher fields ($B_{trans} \gtrsim 5$ G), one observes the appearance of side features positioned symmetrically relative to the $B_{long} = 0$ value and shifting toward higher $B_{long}$ values with increasing $B_{trans}$. In Figure 5(c), we present the shifts of the transition frequencies between the levels $m_s = 0$ and $m_s = \pm 1$ for the four crystallographic axes where the longitudinal field is scanned along [100] in the presence of a transverse field of 13 G at the azimuthal angle of 147.5 degrees[13]. The crossings between transition energies corresponding to different orientations at different values of a longitudinal field match the positions of the measured cross-relaxation features in Figure 5(a). For comparison, see the purple curve, in the presence of a 14.68 G transverse field). In the case of NV center containing nanodiamonds, the application of a transverse magnetic field only reduces the contrast, see Fig. 5(b). This is because we are measuring an ensemble of nanodiamonds with several NVs inside and each one is randomly oriented with respect to the applied magnetic field. Figure 5(d) is the simulation for the shape of the zero-field feature in NDs of random orientations. The graph shows an averaged transition energy difference between states $m_s = 0$ and $m_s = \pm 1$ in the presence of different intensities of a transverse field for many orientations. This simplified model does not account for the transverse strain in diamonds, therefore, yielding narrow curves for the smallest field values.

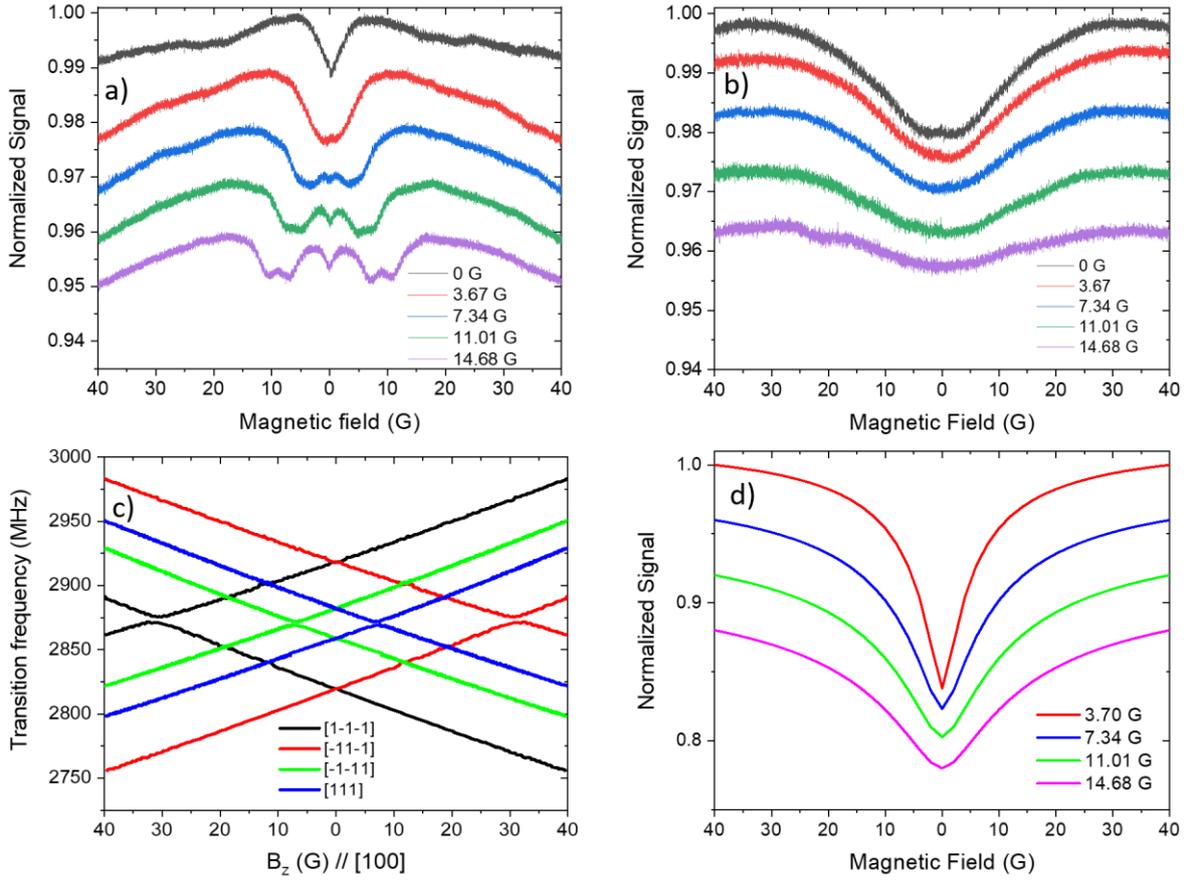

Figure 5. Zero-field resonance in the presence of a transverse field $B_{trans}$ of different intensities for a bulk sample with [100] orientation relative $B_{long}$ (a) and sample with randomly oriented 140 nm nanodiamonds (b). Figure c) shows the calculated transition frequencies for a bulk diamond and $B_{trans}$=14.68 G, and Figure (d) is the simulation of the dip shape for nanodiamonds in a transverse magnetic field of different intensities.

## Discussion

We studied the spin-relaxation time $T_1$ of the ND samples versus the magnetic field and found its strong correlation with the shape of the zero-field resonance. Specifically, the width of the zero-field resonance appears to be a measure of the longitudinal relaxation rate. In a similar measurement with an oriented bulk sample, well-resolved cross-relaxation resonances have been observed. In arbitrary-oriented ND samples, however, all cross-relaxation features are degenerate with respect to crystallographic orientations, which results in a broadening and severe contrast reduction of the zero-field feature. These results underpin the hypothesis that cross-relaxation between nearby NV centers is responsible for the zero-field PL feature. The remarkable agreement between the $T_1$ curves and the PL intensity curves suggests that nothing other than spin relaxation needs to be considered to explain the zero-field feature.

Additionally, we investigated the zero-field microwave-free luminescence feature in various samples of arbitrarily oriented nanodiamonds and bulk diamond plates. First, we considered the excitation-power dependence of our samples and showed that the ND and bulk samples exhibit similar power dependence in both the contrast and the width of the zero-field feature. As mentioned earlier, the decrease of the width and the contrast of the zero-field feature for increasing laser power is explained by the phenomenon called light narrowing, which occurs when the dwell time in the ground state is either comparable to or shorter than the $T_1$ time. Interestingly, we observed an order of magnitude longer $T_1$ times for bulk NV centers than for the nanodiamond samples, yet the power dependence of the zero-field feature shows similar characteristics in the two kinds of samples. Understanding light narrowing in its full depth would require systematic measurement of $T_2^*$ and $T_2$, which is beyond the scope of the present work.

Moreover, we investigated the dependence of various properties of the zero-field feature as a function of the size of the nanodiamond with NV. While the widths and the *E* splittings exhibit a monotonous decrease with the diamond size and approach the values typical for bulk diamonds, the contrast of the zero-field feature exhibits interesting non-monotonous dependence. On the basis of these new results, we may obtain a deeper insight into the root cause of enhanced spin relaxation at zero magnetic field. According to the explanation provided in Ref.[20], there are short-lived NV centers, the so-called fluctuators, that are exposed to charge-state instability and yield unpolarized NV centers. At zero magnetic fields, all orientations are coupled, whereas at nonzero magnetic fields, the different orientations decouple, leading to a change in the effective concentration of the interacting NV centers. In addition to fluctuators, transverse electric fields, and double-flip relaxation processes were also discussed in Ref. [21] as plausible sources of spin relaxation at zero magnetic fields. In addition, the dependence on the angle of the excitation polarization and the NV axis, giving rise to a varying degree of spin polarization among the four NV center orientations, is a plausible mechanism that should be considered[22]. For the sake of clarity, we classify these microscopic explanations as either i) intrinsic or as ii) extrinsic mechanisms. The former category includes models that explain the experiments by using the Hamiltonian of an NV bath (including no fluctuators). Intrinsic mechanisms thus include electric-field-induced relaxation, double-flip processes, and excitation polarization dependence. The fluctuator model deviates from these mechanisms in the sense that other defects, such as nearby P1 centers or the surface, may induce the charge instability of the NV center and cause the zero-field feature. Thus, this model includes extrinsic objects that couple to the NV bath.

Our ODMR results show that the effective zero-field splitting of the spin sublevels of the NV centers in ND increases with decreasing ND diameters, indicating an increase of the strain and/or electric-field noise. As a consequence of the wider distribution of the NV energy levels, one would expect a broadening of the zero-field feature, which is confirmed by our measurements. Considering strain and electric field fluctuation as a possible source of the zero-field feature, one would expect a sharp increase of the contrast for decreasing the ND size below 500 nm. The opposite is observed in our measurement. Furthermore, the decrease in the contrast when going from 750-nm NDs to 3000-nm NDs cannot be explained by electric field and strain, as the *E* splitting is constant in this interval. Therefore, our findings do not support strain- and electric-field-related mechanisms as the dominant sources of zero-field relaxation. Double-flip relaxation processes originate from the dipole-dipole coupling, for which we do not expect dependence on the size of the NDs as long as the NDs have the same concentration of NV centers. As can be seen in Table 1, this condition is approximately met, thus the relevance of double-flip relaxation processes is not confirmed with our nanodiamond samples. Similarly, we do not expect any nanodiamond-size dependence of the polarization effects.

Finally, we discuss the viability of the fluctuator model. It is well-known that the stability of NV centers degrades near the surface. In particular, through the effect of band bending and the abundance of surface defects, the surface can have a significant impact on the charge stability of the NV centers and thus on the concentration of fluctuators. In fact, band bending can be detected as far as ~40 nm from the surface[23,24], which is comparable with the radius of our smallest nanodiamonds. We also mention here that, as expected, our nanodiamonds are not spherically symmetric, therefore, their surface may include facets and sharp edges, which increases the volume affected by band bending and surface defects. Therefore, we anticipate that the number of fluctuators increases near the surface of the nanodiamonds, which, in turn, can enhance relaxation effects and increase the contrast of the zero-field feature. In this respect, the surface-volume ratio may play an important role for NDs. For decreasing ND size, the surface-volume ratio increases and thus one can expect an increasing contrast. A similar trend is observed between 500 nm and 3000 nm in our measurements, although, we have low sampling in this region. Below 500 nm, the contrast decreases sharply. We attribute this effect to the increased spin relaxation at the surface and the overwhelming increase of the fluctuator in small nanodiamonds. The total T$_1$ time of the NV ensemble is obtained as $T_1^{-1} = (T_1')^{-1} + (T_1^{NV-bath})^{-1}$, where $T_1'$ includes relaxation effects due to phonons, charge fluctuations, and other spins in the sample, while $T_1^{NV-bath}$ accounts solely for the contributions due to the coupling between the NV centers. As $T_1'$ decreases, for example, due to the presence of the surface and the increase of the number of short-lived NV centers, the $T_1^{NV-bath}$ term may become smaller than to $T_1'$. In this case, the contrast of the zero-field feature reduces. In conclusion, to interpret our observations in nanodiamonds, we needed to introduce other defect types through the fluctuator model.

**Conclusions**

We studied the spin-relaxation time $T_1$ of the ND samples versus the magnetic field as a function of the size of the nanodiamonds. We found evidence that cross-relaxation occurs between nearby NV centers in the nanodiamonds, which results in a well-observable change in the fluorescence intensity at near-zero magnetic fields. The reported results can be used in scalar and vector magnetometry applications and constitute the basis of a practical microwave-free or even all-optical technique. The fact that the zero-field feature does not depend on the ND orientation suggests a particularly straightforward modality of magnetometry and magnetic imaging which will be reported in more detail in a subsequent publication. It relies on covering a magnetic surface to be examined/inspected with an ND layer of arbitrary shape and orientation (like with a salt/pepper dispenser) and directly imaged with a wide-field setup. We call this modality a "salt-and-pepper" technique.

**Acknowledgments**

The authors thank Klaudia Kvakova and Petr Cigler for providing nanodiamonds with nitrogen-vacancy color centers. The preparation of nanodiamond samples was supported by the Czech Academy of Sciences – Strategy AV21 (VP29). M.M., A.M.W. and W.G. acknowledge support from the TEAM NET programme of the Foundation for Polish Science co-financed by the European Union under the European Regional Development Fund, project POIR.04.04.00-00-1644/18. This research was also funded in part by the National Science Centre, Poland grant number 2020/39/I/ST3/02322. D.B., O.D., A.W., and T.L. acknowledge support from the European Commission's Horizon Europe Framework Program under the Research and Innovation Action MUQUABIS GA no. 101070546, by the German DFG, Project SFB 1552 "Defekte und Defektkontrolle in weicher Materie" and funding by the Carl-Zeiss-Stiftung (HYMMS P2022-03-044), also the German Federal Ministry of Education and Research (BMBF) within the Quantumtechnologien program through the DIAQNOS project (project no. 13N16455). A.G. acknowledges Horizon Europe projects QuMicro (Grant No. 101046911) and SPINUS (Grant No. 101135699) also the National Research, Development and Innovation Office of Hungary (NKFIH) within the Quantum Information National Laboratory of Hungary (Grant No. 2022-2.1.1-NL-2022-00004) and within project FK 145395. V.I. acknowledges the support of the Knut and Alice Wallenberg Foundation through the WBSQD2 project (Grant No. 2018.0071). The authors thank Shimon Kolkowitz for fruitful discussions.